\documentclass[preprint,showpacs,preprintnumbers,amsmath,amssymb]{revtex4}
\usepackage{amsfonts}
\usepackage{amssymb}
\usepackage{latexsym}
\usepackage{graphicx}

\newcommand{\al}{\alpha}
\newcommand{\be}{\beta}
\newcommand{\pa}{\partial}

\newcommand{\na}{\nabla}

\newcommand{\rar}{\rightarrow}

\begin{document}

\preprint{M\'exico ICN-UNAM 04/02, \  March, 2002}
\title{$H_3^{++}$ molecular ion in a strong magnetic field:
triangular configuration\\ {\it Physical Review A (accepted)}}
\author{J. C. \surname{L\'opez Vieyra}}
\email{vieyra@nuclecu.unam.mx}
\affiliation{}
\author{A. V. Turbiner}
\altaffiliation[]{On leave of absence from the Institute for Theoretical
 and Experimental Physics, Moscow 117259, Russia}
\email{turbiner@nuclecu.unam.mx}
\affiliation{Instituto de Ciencias Nucleares, UNAM,
Apartado Postal 70-543, 04510 M\'exico}

\begin{abstract}
  The existence of the molecular ion $H_3^{++}$ in a magnetic field in
  a triangular configuration is revised. A variational method with an
  optimization of the form of the vector potential (gauge fixing) is
  used.  It is shown that in the range of magnetic fields $10^8 < B <
  10^{11}\,G$ the system $(pppe)$, with the protons forming an
  equilateral triangle perpendicular to the magnetic line, has a
  well-pronounced minimum in the total energy. This configuration is
  unstable under the decays $H\mbox{-atom} + p + p$ and $H_2^+ + p$.
  The triangular configuration of $H_3^{++}$ complements $H_3^{++}$ in
  the linear configuration which exists for $B \gtrsim 10^{10}\,G$.
\end{abstract}


\maketitle

\section{Introduction}
\label{sec:intro}

Recently, it was announced that the molecular ion $H_3^{++}$ in
a linear configuration can exist in a strong magnetic field $B
\gtrsim 10^{10}\,G$ and become even the most stable one-electron
system at $B \gtrsim 10^{13}\,G$ \cite{Turbiner:1999,
Lopez-Tur:2000}. The goal of this article is to study whether the
$H_3^{++}$ molecular ion can exist in a certain spatial
configuration -- the protons form an equilateral triangle while a
magnetic field is directed perpendicular to it. This configuration
was already studied in \cite{Lopez:2000} with an affirmative answer.
In the present work we will show that an improper gauge dependence of
the trial functions in \cite{Lopez:2000} caused a significant loss of
accuracy and led to qualitatively incorrect results.

The Hamiltonian which describes three infinitely heavy protons and
one electron placed in a uniform constant magnetic field directed
along the $z-$axis, ${\bf B}=(0,0,B)$ is given by
\begin{equation}
\label{Ham}
 {\cal H} = {\hat p}^2 + \frac{2}{R_{ab}}+ \frac{2}{R_{ac}}+
 \frac{2}{R_{bc}} -\frac{2}{r_1} -\frac{2}{r_2}- \frac{2}{r_3}  +
2({\hat p} {\cal A}) +  {\cal A}^2 \ ,
\end{equation}
(see Fig.1 for notations), where ${\hat p}=-i \na$ is the
momentum, ${\cal A}$ is a vector potential, which corresponds to
the magnetic field $\bf B$. We assume that the protons $a,b,c$
form an equilateral triangle, $R_{ab}=R_{bc}=R_{ac}=R$, and the
magnetic field $\bf B$ is directed perpendicular to it. This
configuration of the protons is stable from classical-mechanical
point of view, since electrostatic repulsion of the protons is
compensated by the Lorentz force. It justifies more the use of the
Born-Oppenheimer approximation and also adds extra stability to
the whole system $(pppe)$. A small perturbation of a proton
position directed outside the plane perpendicular to the magnetic
line can distort the above triangular configuration. However, our
calculations show that the presence of the electron can stabilize
the configuration, at least, for small perturbations. Thus, the
stability of this configuration is of a different nature than the
linear one. There it appears to be a consequence of the
quasi-one-dimensionality of the problem and the compensation of
the proton repulsion by the interaction with one-dimensional
electronic cloud \cite{Turbiner:1999, Lopez-Tur:2000}.

\begin{figure}[tb]
\begin{center}
     \includegraphics*[width=3.2in,angle=-90]{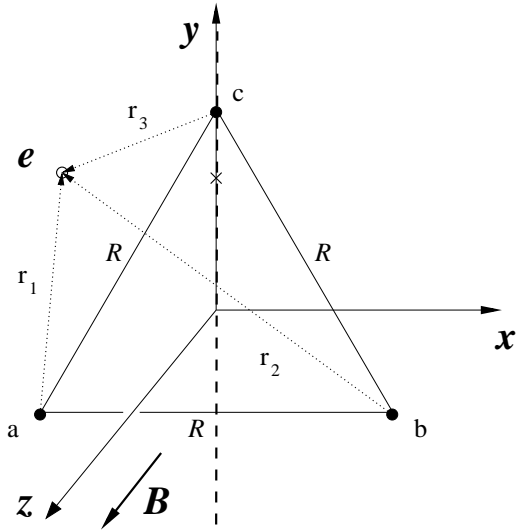}
     \caption{Geometrical setting  for the $H_3^{++}$ ion in a
      magnetic field directed along the $z$-axis. The protons are
      marked by bullets and are situated in the $x-y$ plane. It is
      assumed that the gauge center is situated
      on bold-dashed line which connects the center of the triangle
      and the position of the proton $c$ (see text).}
    \label{fig:1}
\end{center}
\end{figure}

Atomic units are used throughout ($\hbar$=$m_e$=$e$=1) albeit
energies are expressed in Rydbergs (Ry). Sometimes, the magnetic
field $B$ is given in a.u. with $B_0= 2.3505 \times 10^9 G$.

\section{Optimization of vector potential}

It is well known that the vector potential for a given magnetic
field, even taken in the Coulomb gauge $(\na \cdot {\cal A}) =0$,
is defined ambiguously, up to a gradient of an arbitrary function.
This gives rise a feature of gauge invariance: the Hermitian
Hamiltonian is gauge-independent as well as the eigenvalues and
other observables. However, since we are going to use an
approximate method for solving the Schroedinger equation with the
Hamiltonian (\ref{Ham}), our energies can be gauge-dependent (only
the exact ones should be gauge-independent). Hence one can choose
the form of the vector potential in a certain optimal way, looking
for a gauge which leads to minimal energy for given class of trial
functions. In particular, if the variational method is used an
optimal gauge can be considered as a variational function and then
is chosen by a procedure of minimization.

Let us consider a certain one-parameter family of vector
potentials corresponding to a constant magnetic field $B$ (see,
for example, \cite{Larsen})
\begin{equation}
\label{Vec}
  {\cal A}= B(-\xi(y-y_0), (1-\xi) (x-x_0), 0)\ ,
\end{equation}
where $\xi, x_0, y_0$ are parameters. The position of the {\it gauge
  center}, where ${\cal A}(x,y)=0$, is defined by $x=x_0, y=y_0$. If
the gauge center is at the origin, $x_0=y_0=0$, and $\xi=1/2$ we
get the well-known and widely used gauge which is called symmetric
or circular. If $\xi=1$, it corresponds to the asymmetric or
Landau gauge (see \cite{LL}). By substituting (\ref{Vec}) into
(\ref{Ham}) we arrive at the Hamiltonian in the form

\begin{widetext}
\begin{eqnarray}
  \label{Ham.fin}
 {\cal H} &=& -{\nabla}^2 + \frac{6}{ R} -\frac{2}{r_1}
 -\frac{2}{r_2}-\frac{2}{r_3}
 +  2 i B[-\xi(y-y_0)\pa_x  + (1-\xi) (x-x_0)\pa_y]  \nonumber \\
& & +  B^2 [ (1-\xi)^2 (x-x_0)^2+ \xi^2 (y-y_0)^2] \ ,
\end{eqnarray}
\end{widetext}
where $R$ is the size of the triangle side.

The idea of choosing an optimal (convenient) gauge has widely been
exploited in quantum field theory calculations. It has also been
discussed in quantum mechanics, in particular, in connection with
the present problem (see, for instance, \cite{Schmelcher} and
references therein). Perhaps, the first constructive (and
remarkable) attempt to apply this idea was made by Larsen
\cite{Larsen}. In his variational study of the ground state of the
$H_2^+$ molecular ion it was explicitly shown that gauge
dependence on energy can be quite significant. Furthermore even an
oversimplified optimization procedure improves the numerical
results.

Our present aim is to study the ground state of (\ref{Ham}),
(\ref{Ham.fin}). It can be easily demonstrated that for a one-electron
problem there always exists a certain gauge for which the ground state
eigenfunction is a real function. Let us fix a vector potential in
(\ref{Ham}). Assume that we have solved the spectral problem exactly and
have found the exact ground state eigenfunction. In general, it is a
certain complex function with a non-trivial, coordinate-dependent phase.
Considering their phase as gauge phase and then gauging it away,
finally, it will result in a new vector potential. This vector
potential has the property we want -- the ground state eigenfunction
of the Hamiltonian (\ref{Ham}) is real. It is obvious that similar
considerations can be performed for any excited state. In general, for
a given eigenstate there exists a certain gauge in which the
eigenfunction is real. These gauges can be different for different
eigenstates. A similar situation takes place for any one-electron
problem.

Dealing with real trial functions has an obvious advantage: the
expectation value of the term $\sim {\cal A}$ in (\ref{Ham}) or $\sim
B$ in (\ref{Ham.fin}) vanishes when is taken over any real,
normalizable function. Thus, without loss of generality, the term
$\sim B$ in (\ref{Ham.fin}) can be omitted. Furthermore, it can be
easily shown that, if the original problem possesses axial symmetry
with axis coinciding with the direction of the magnetic field, the
real ground state eigenfunction always corresponds to the symmetric
gauge.

\section{Choosing trial functions}

The choice of trial functions contains two important ingredients:
(i) a search for the gauge leading to the real ground state
eigenfunction and (ii) performance of a variational calculation
based on {\it real} trial functions. The main assumption is that a
gauge corresponding to a real ground state eigenfunction is of the
type (\ref{Vec}) (or somehow is close to it)\footnote{It can be
formulated as a problem
--
  for a fixed value of $B$ and a fixed size of triangle, to find a
  gauge for which the ground state eigenfunction is real.}. In other
words, one can say that we look for a gauge of the type (\ref{Vec})
which admits the best possible approximation of the ground state
eigenfunction by real functions. Finally, in regard to our problem the
following recipe of variational study is used: \emph{First of all, we
  construct an {\bf adequate} variational real trial function
  \cite{Tur}, which reproduces the original potential near Coulomb
  singularities and at large distances, where $\xi, x_0, y_0$ would
  appear as parameters.  Then we perform a minimization of the energy
  functional by treating the trial function's free parameters and
  $\xi, x_0, y_0$ on the same footing.}  In particular, such an
approach enables us to find eventually the \emph{optimal} form of the
Hamiltonian as a function of $\xi, x_0, y_0$. It is evident that for
small interproton distances $R$ the electron prefers to be near the
center of the triangle (coherent interaction with all three protons),
hence $x_0, y_0$ should correspond to the center of the triangle. In
the opposite limit of large $R$ the electron is situated near one of
the protons ( a situation of incoherence - the electron selects and
then interacts essentially with one proton), therefore $x_0, y_0$
should correspond to the position of a proton. We make a natural
assumption that the gauge center is situated on a line connecting the
center of the triangle and one of the protons, hence $$x_0=0\ ,\
y_0=\frac{R}{\sqrt{3}}d\ ,$$
(see Fig.1). Thus, the position of the
gauge center is measured by the parameter $d$ -- the relative distance
between the center of triangle and the gauge center. If the gauge
center coincides with the center of the triangle, then $d=0$. On the
other hand, if the gauge center coincides with the position of proton,
$d=1$.

The above recipe was successfully applied in a study of the
$H_2^+$-ion in a magnetic field \cite{Lopez:1997} and led to
prediction of the existence of the exotic ion $H_3^{++}$ at $B
\gtrsim 10^{10}\,G$ in a linear configuration \cite{Turbiner:1999,
Lopez-Tur:2000}.

One of the simplest trial functions satisfying the above-mentioned
criterion is
 \begin{equation}
 \label{tr:1}
 \Psi_1= {e}^{-\al_1  (r_1+r_2+r_3)
  - B  [\be_{1x} (1-\xi) (x-x_0)^2 + \be_{1y}\xi (y-y_0)^2] }\ ,
 \end{equation}
(cf. \cite{Lopez:1997}), where $\al_1,\be_{1x,1y},\xi, x_0, y_0$
are variational parameters. The requirement of normalizability of
(\ref{tr:1}) implies that $\al_1,\be_{1x,1y}$ are non-negative
numbers and $\xi \in [0,1]$. Actually, this is a Heitler-London
type function multiplied by the lowest (shifted) Landau orbital
associated with the gauge (\ref{Vec}). It is natural to assume
that the function (\ref{tr:1}) describes the domain of coherence -
small interproton distances and probably distances near the
equilibrium. Another trial function

\begin{widetext}
 \begin{equation}
 \label{tr:2}
 \Psi_2 = \bigg({e}^{-\al_2 r_1}+ {e}^{-\al_2 r_2}+ {e}^{-\al_2  r_3}\bigg)
{e}^{ -  B  [\be_{2x}(1-\xi) (x-x_0)^2 +\be_{2y}\xi (y-y_0)^2] }\
,
 \end{equation}
\end{widetext}
 (cf. \cite{Lopez:1997}), is of the Hund-Mulliken type multiplied by
 the lowest (shifted) Landau orbital. Here $\al_2, \be_{2x,2y} $,
 $\xi, x_0, y_0$ are variational parameters.  Presumably this function
 dominates for  sufficiently large interproton distances $R$ giving
 an essential contribution there.  Hence, it models an interaction of
 a hydrogen atom and protons (charged centers) and can also describe
 a possible decay mode into them, $H_3^{++} \rar H+p+p$. In a similar
 way one can construct a trial function which would model the
 interaction $H_2^+ + p$,

\begin{widetext}
 \begin{equation}
 \label{tr:3}
 \Psi_3 = \bigg({e}^{-\al_3 (r_1+ r_2)}+ {e}^{-\al_3 (r_1+ r_3)}+
  {e}^{-\al_3 (r_2+ r_3)}\bigg)
  {e}^{ -  B  [\be_{3x} (1-\xi) (x-x_0)^2 +\be_{3y}\xi (y-y_0)^2] }\ .
 \end{equation}
\end{widetext}
One can say that this  is a mixed Hund-Mulliken and Heitler-London type
trial function multiplied by the lowest (shifted) Landau orbital.
Here $\al_3,\be_{3x,3y}$, $\xi, x_0, y_0$ are variational parameters.
It is clear that this function gives a subdominant contribution at
large $R$ and a certain, sizable contribution to a domain of
intermediate distances.

There are two natural ways - linear and non-linear - to
incorporate the behavior of the system both near equilibrium and
at large distances in a single trial function. A general non-linear
interpolation involving the above trial functions is of the form
\begin{widetext}
\begin{eqnarray}
  \label{tr:4-1}
    \Psi_{4-1}&=& \bigg(
   {e}^{-\al_4 r_1-\al_5 r_2-\al_6 r_3}+
   {e}^{-\al_4 r_1-\al_5 r_3-\al_6 r_2}+
   {e}^{-\al_4 r_2-\al_5 r_1-\al_6 r_3}+
   {e}^{-\al_4 r_2-\al_5 r_3-\al_6 r_1}+
\nonumber \\
& &   {e}^{-\al_4 r_3-\al_5 r_1-\al_6 r_2}+
   {e}^{-\al_4 r_3-\al_5 r_2-\al_6 r_1}
  \bigg)  {e}^{ - B  [\be_{4x}(1-\xi) (x-x_0)^2 +\be_{4y}\xi (y-y_0)^2] }
\end{eqnarray}
\end{widetext}
(cf. \cite{Lopez:1997}), where $\al_{4,5,6},\be_{4x,4y},\xi, x_0,
y_0$ are variational parameters. In fact, this is a
Guillemin-Zener type function multiplied by the lowest (shifted)
Landau orbital. If $\al_{4}=\al_{5}=\al_{6}$, the function
(\ref{tr:4-1}) reproduces (\ref{tr:1}). While if
$\al_{5}=\al_{6}=0$, it reproduces (\ref{tr:2}). If
$\al_{4}=\al_{5}$ and $\al_{6}=0$, it reproduces (\ref{tr:3}). The
linear superposition of (\ref{tr:1}), (\ref{tr:2}), (\ref{tr:3})
leads to
 \begin{equation}
 \label{tr:4-2}
 \Psi_{4-2}= A_1 \Psi_{1} + A_2 \Psi_{2}+ A_3 \Psi_{3}\ ,
 \end{equation}
where one of the parameters $A_{1,2,3}$ is kept fixed, being
related to the normalization factor. The final form of the trial
function is a linear superposition of functions (\ref{tr:4-1}) and
(\ref{tr:4-2})
 \begin{equation}
 \label{trial}
 \Psi_{trial}= A_1 \Psi_{1} + A_2 \Psi_{2}+A_3 \Psi_{3}+
 A_{4-1} \Psi_{4-1}\ ,
 \end{equation}
where three out of four parameters $A$'s are defined
variationally. For a given magnetic field the total number of
variational parameters in (\ref{trial}) is 20, when $\xi$ and $d$
are included. Calculations were performed  using the minimization
package MINUIT of CERN-LIB. Numerical integrations were carried
out with relative accuracy $\sim 10^{-7}$ by use of the adaptive
NAG-LIB (D01FCF) routine. All calculations were performed on a PC
Pentium-II $450 MHz$.

\section{Results}

Our variational study shows that in the range of magnetic fields
$10^8 < B < 10^{11}\,G$ the system $(pppe)$, with the protons
forming an equilateral triangle perpendicular to the magnetic
line, has a well-pronounced minimum in the total energy (see Table
1 and Fig. 2-5). With a magnetic field increase the total energy
gets larger and the size of triangle shrinks but the height of the
barrier increases (for example, it grows from $\sim 0.028$ Ry at
$10^9$ G to $\sim 0.037$ Ry at $10^{10}$ G). It was checked that
the equilibrium configuration remains stable under small
deviations of the proton positions but is unstable globally,
decaying to $H + p + p$ and $H_2^+ + p$. This implies the
existence of the molecular ion $H_3^{++}$ in a triangular
configuration for the range of magnetic fields $10^8 < B <
10^{11}\,G$.

\begingroup
\squeezetable
\begin{table}
\begin{ruledtabular}
\begin{tabular}{clcccc}
& & & & & \\ \( B \) (G)& & \( H_{3}^{++} \)(triangle)
 & \(H_{3}^{++} \)(linear)& \( H \)-atom & \( H_{2}^{+} \)
 (parallel) \\[9pt]  \hline
 & & & & & \\ & \( E_{T} \) (Ry) & -0.524934 & --- & -0.920821
 &-1.150697\\[5pt] \( 10^{9} \) &\( R \) (a.u.)    & 3.161  & &
 &1.9235\\[5pt] &\( \xi \) & 0.50005 & & &
 \\[5pt] &\( d \) & 0.0 & &             &
 \\[5pt]
\hline  & & & & & \\ &\( E_{T} \)  (Ry) & 2.724209
 & 1.846367 & 1.640404 & 1.090440\\[5pt] \( 10^{10} \) &\( R \)
 (a.u.)  &1.4012  & 2.0529 &            & 1.2463\\[5pt] &\( \xi \)
 &0.50102 &            & &
 \\[5pt] &\( d \)           &0.00041 &            & &
 \\[5pt]
\hline  & & & & & \\ &\( E_{T} \)  (Ry) & 19.331448
 & 16.661543 & 16.749684 & 15.522816 \\[5pt] \(5\ 10^{10} \) &\( R \)
 (a.u.)  &0.7766  & 1.0473 &            & 0.7468\\[5pt] &\( \xi \)
 &0.50205 &            & &
 \\[5pt] &\( d \)           &0.0011 &            & &
\end{tabular}
\end{ruledtabular}
\caption{Total energy, equilibrium distances and characteristics
of the vector potential (\ref{Vec}). Comparison with $H_3^{++}$ in
a linear configuration aligned along the magnetic line
\cite{Lopez-Tur:2000, Lopez:1997},  hydrogen atom
\cite{Potekhin:2001} as well as the $H_2^+$-ion aligned along the
magnetic line \cite{Lopez-Tur:2000, Lopez:1997} is given.}
\end{table}
\endgroup

\begin{figure}[htbp]
\begin{center}
   {\includegraphics[width=2.5in,angle=-90]{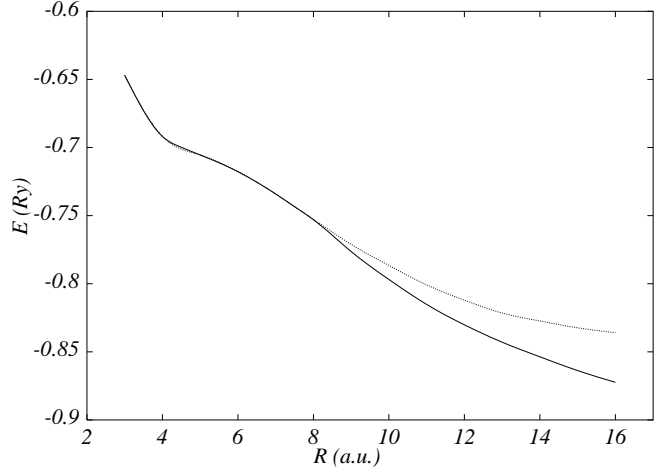}}
   \caption{Total energy of $(pppe)$ at $10^8$G as function of
     the size of the triangle (solid curve). The dotted line
     is a result of minimization if $d=0$ (the gauge center
     coincides with the center of the triangle).}
   \label{fig:2}
\end{center}
\end{figure}

\begin{figure}[htbp]
\begin{center}
   {\includegraphics*[width=2.5in,angle=-90]{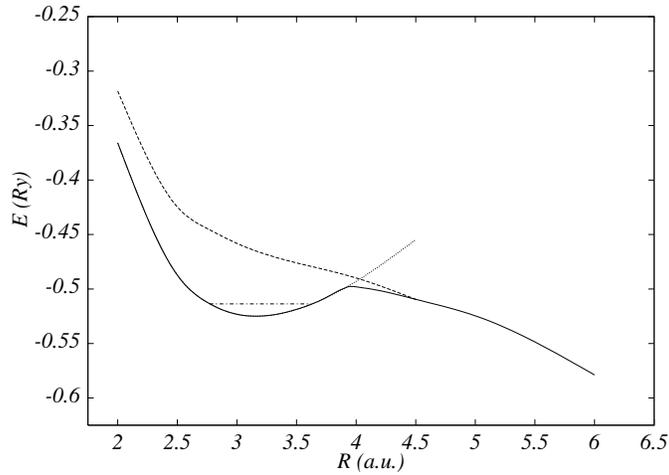}}
   \caption{Total energy of $(pppe)$ at $10^{9}$G as function of
     the size of the triangle
     (solid line). The dotted line is the result of minimization if
     $d=0$ is kept fixed.  The dashed line describes a
     result of minimization if $d=1$ (the gauge center and position
     of a proton coincide, see text). The dot-dashed line displays
     the position of the first vibrational state.}
   \label{fig:3}
\end{center}
\end{figure}

\begin{figure}[htbp]
\begin{center}
  { \includegraphics*[width=2.5in,angle=-90]{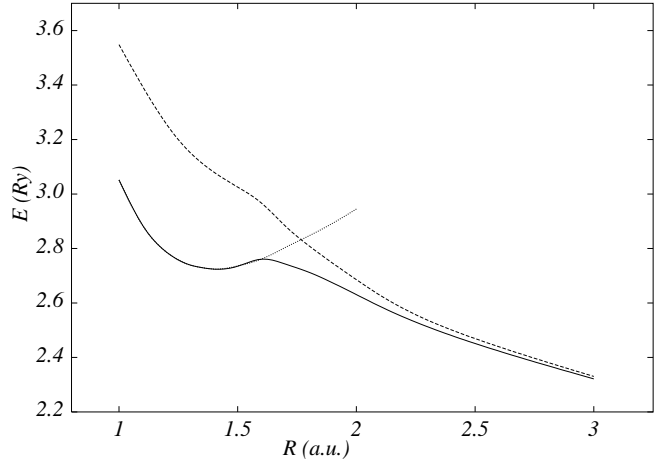}}
   \caption{Total energy of the system $(pppe)$ at $10^{10}$G as
   function of the size of the triangle (solid line). The dotted
   line is the result of minimization if $d=0$ are kept fixed.
   The dashed line describes a result
   of minimization if $d=1$ (the gauge center and position
   of proton coincide, see text).}
   \label{fig:4}
\end{center}
\end{figure}

\begin{figure}[htbp]
\begin{center}
   {\includegraphics*[width=2.5in,angle=-90]{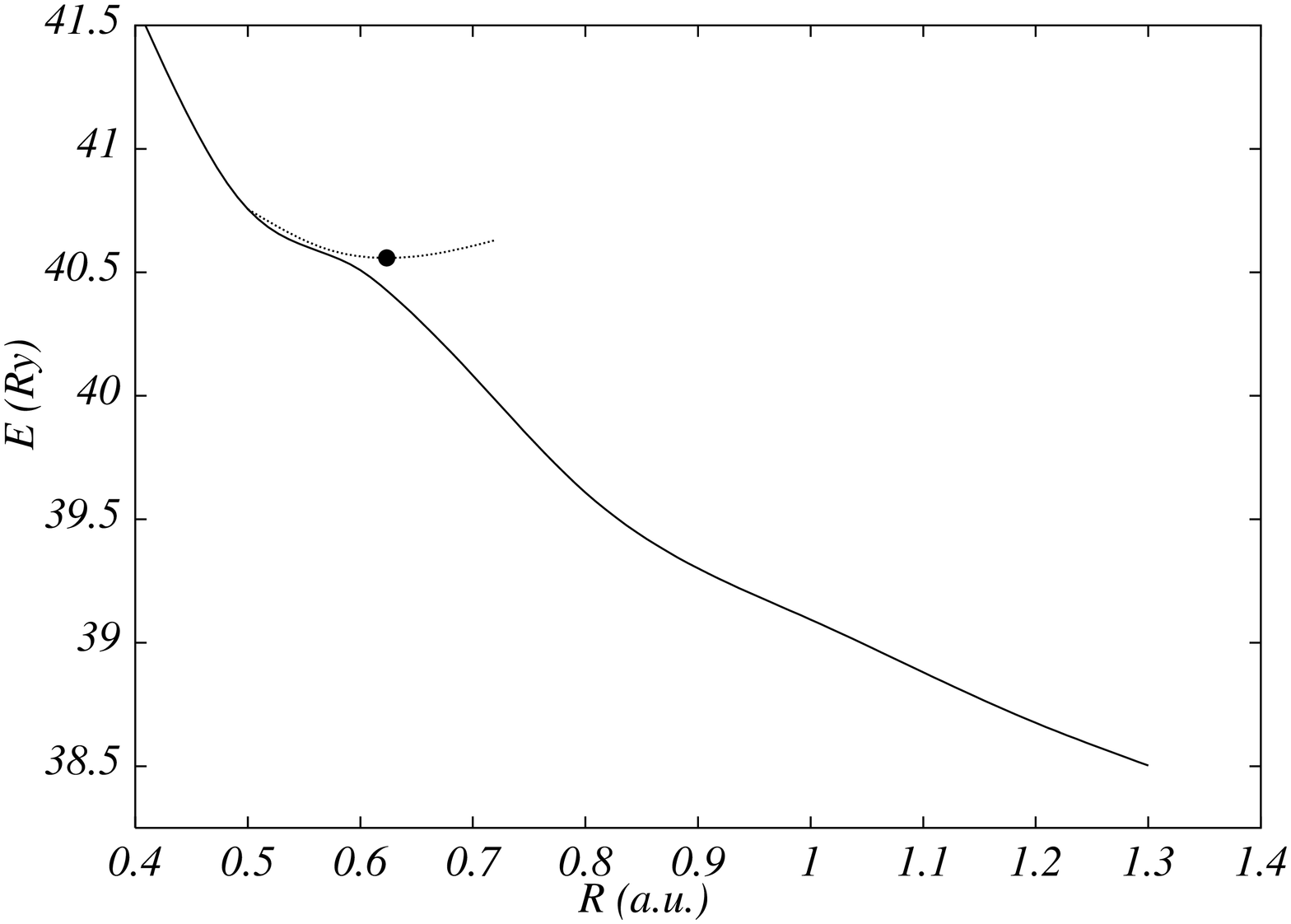}}
   \caption{Total energy of the system $(pppe)$ at $10^{11}$G as
   function of  the size of the triangle (solid line).
    The bullet denotes the position of a spurious minimum which
    appear if gauge center is kept fixed at $x_0=y_0=0$\ ($d=0$,
    dotted line)
    (the gauge center and the center of the triangle coincide,
    see \cite{Lopez:2000}).}
   \label{fig:5}
\end{center}
\end{figure}

Our calculations show that the equilibrium position always
corresponds to the situation when the gauge center coincides with
the center of the triangle, $d=0$. Therefore, the optimal vector
potential appears in the symmetric gauge, $\xi=0.5$ (see Table 1
and discussion above). In Figures 2 and 5 two typical situations
of absence of a bound state are presented. At $B=10^8$ G a certain
irregularity appears on the potential curve but neither $d=1$ and
$d=0$, nor $d_{min}$ curves develop a minimum. A similar situation
holds for smaller magnetic fields $B < 10^8$. At $B=10^{11}$ G the
situation is more complicated. If the gauge center is kept fixed
and coincides with the center of the triangle, the potential curve
displays a very explicit minimum, which disappear after varying
the gauge center position (!). Something analogous to what is
displayed in Fig. 5 appears for larger magnetic fields, $B >
10^{11}$ G. This artifact of the gauge center fixing at $d=0$ had
led to an erroneous statement in \cite{Lopez:2000} about the
existence of $H_3^{++}$ in a triangular configuration at $B \geq
10^{11}$ G.

Fig. 3 displays the plots of different potential curves
corresponding to the gauge center fixed at the position of one
proton, at the center of the triangle and also varying the gauge
center at $B = 10^{9}$ G. A curve describing the total energy
demonstrates a clear, sufficiently deep minimum. As was expected
small distances correspond to a gauge center coinciding with the
center of the triangle, while large distances are described by a
gauge center situated on a proton. It is important to emphasize
that the domain of near-equilibrium distances (and approximately
up to the position of the maximum) is described by the
gauge-center-on-center-of-triangle curve.  The well keeps a
vibrational state with energy $E_{vib} = 0.0112654$ Ry. In Fig.4
there are plots of different potential curves corresponding to the
gauge center fixed at the position of one proton, at the center of
the triangle and also varying the gauge center at $B = 10^{10}$ G.
A curve describing the total energy demonstrates a clear,
sufficiently deep minimum. Unlike the situation for $B = 10^{9}$
G, this well is unable to keep a vibrational state.  Similar to
what happens for $B = 10^{9}$ G, small distances correspond to a
gauge center coinciding with the center of the triangle, large
distances are described by a gauge center situated on a proton,
the domain of near-equilibrium distances and up to the position of
the maximum is described by the gauge-center-on-center-of-triangle
curve. It is quite interesting to investigate the behaviour of the
gauge center position $d$ as well as a gauge "asymmetry", $\xi$
versus $R$.  Both plots are of a phase transition-type, with
change of behavior near the maximum of the barrier (see Figs.
6-7). The width of the transition domain is $\sim 0.02~a.u.$ (and
$\sim 0.1~a.u.$ for $B = 10^{10}$ G). The evolution of the
electronic distributions with respect to the size of the triangle
is shown on Figs.8-9 for $10^{9}$ G and $10^{10}$ G, respectively.
For small and intermediate $R$ at $B=10^{9}$ G the distribution is
characterized by three more or less equal peaks corresponding to
the proton positions. However, it changes drastically after
crossing the point of phase transition at $R \sim 3.93$ a.u. One
peak disappears almost completely, while another one reduces its
height.  At large distances two peaks disappear completely, the
distribution is characterized by one single peak, centered
approximately at the position of one of the protons. For the case
of $B=10^{10}$ G the electronic distribution is always
characterized by a single peak, which is situated at the center of
the triangle at small and intermediate distances. Then at $R >
1.7$ a.u.  the position of the peak shifts to a position of the
proton. For both values of the magnetic field at asymptotically
large distances the center of the peak coincides exactly with the
position of the proton.

\begin{figure}[htbp]
\begin{center}
    \includegraphics*[width=2.5in,angle=-90]{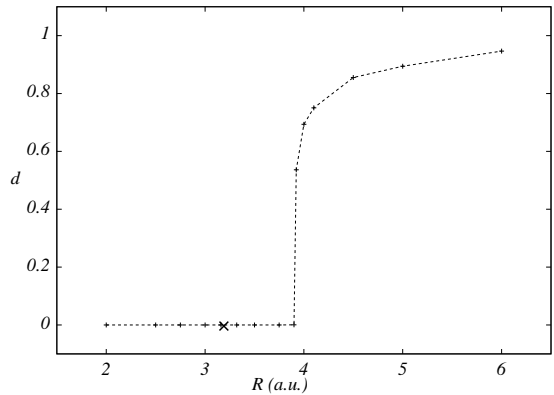}
    \caption{Dependence of the position of the gauge
      center $d$ on the size of the triangle for $10^9$ G.}
   \label{fig:6}
\end{center}
\end{figure}

\begin{figure}[htbp]
\begin{center}
    \includegraphics*[width=2.5in,angle=-90]{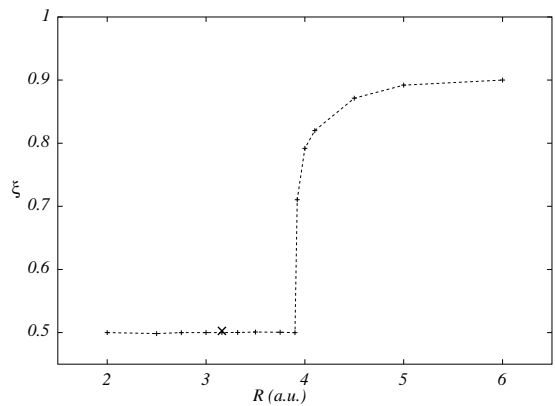}
    \caption{Dependence of the parameter $\xi$ on the size of
    the triangle for $10^9$ G.}
   \label{fig:7}
\end{center}
\end{figure}

\begin{figure*}
\[
\begin{array}{ccc}
\begin{picture}(200,150)(0,0)
\put(0,150){\includegraphics[width=120pt,angle=-90]{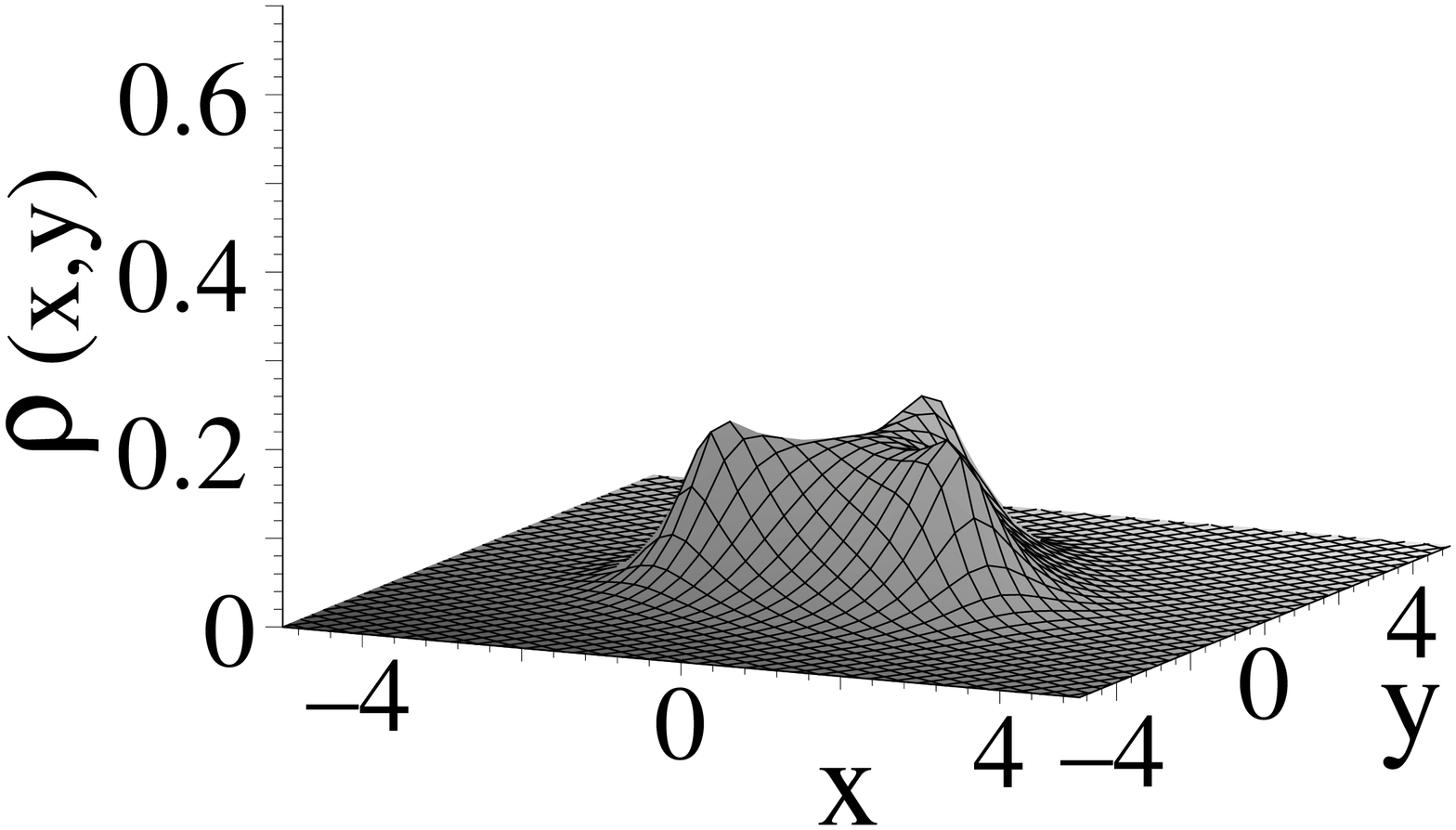}}
\put(120,100){$R=R_{eq}$}
\end{picture}
&  &
\begin{picture}(200,150)(0,0)
\put(0,150){\includegraphics[height=120pt,width=120pt,angle=-90]{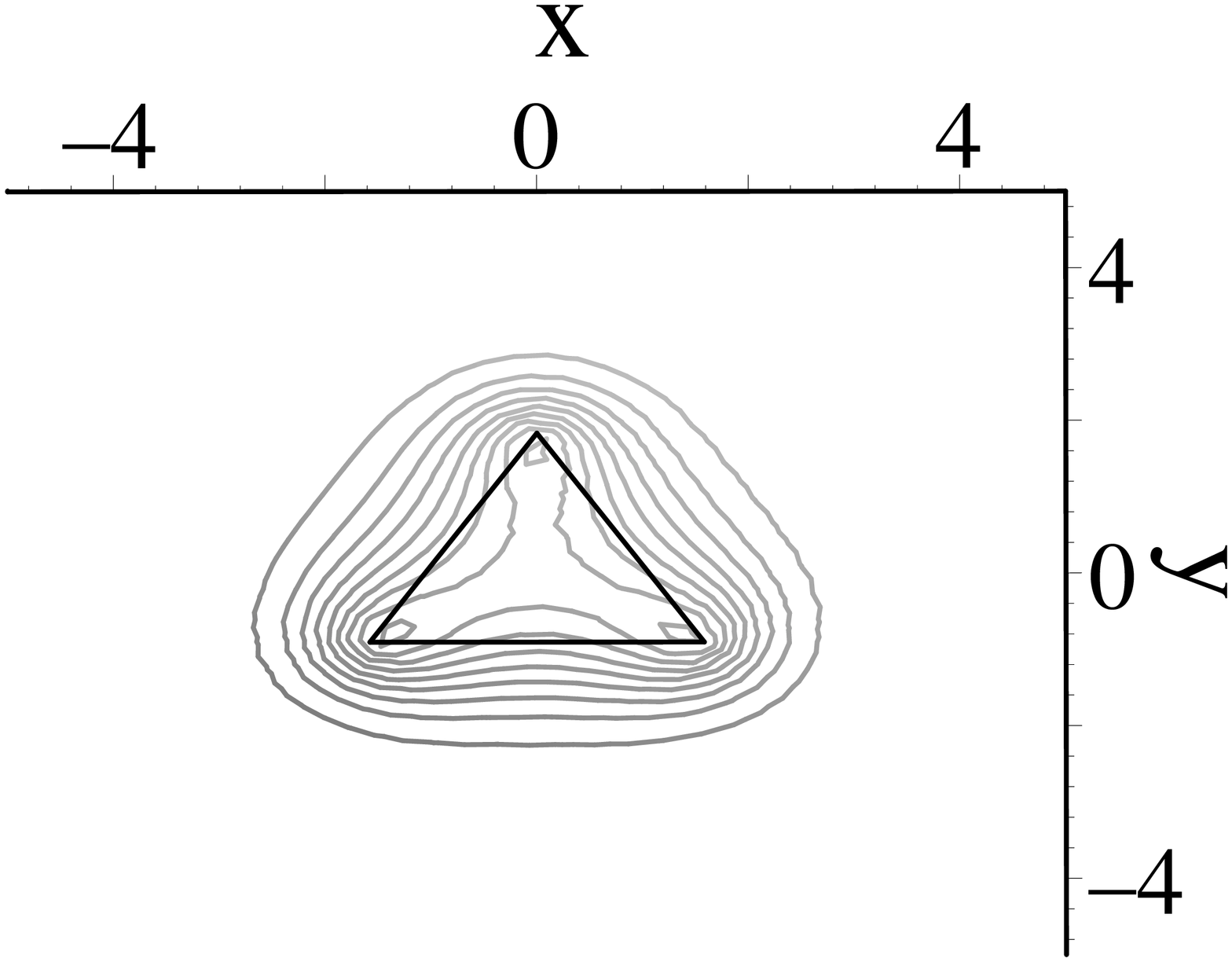}}
\end{picture}
\\[-10pt]
\begin{picture}(200,150)(0,0)
\put(0,150){\includegraphics[width=120pt,angle=-90]{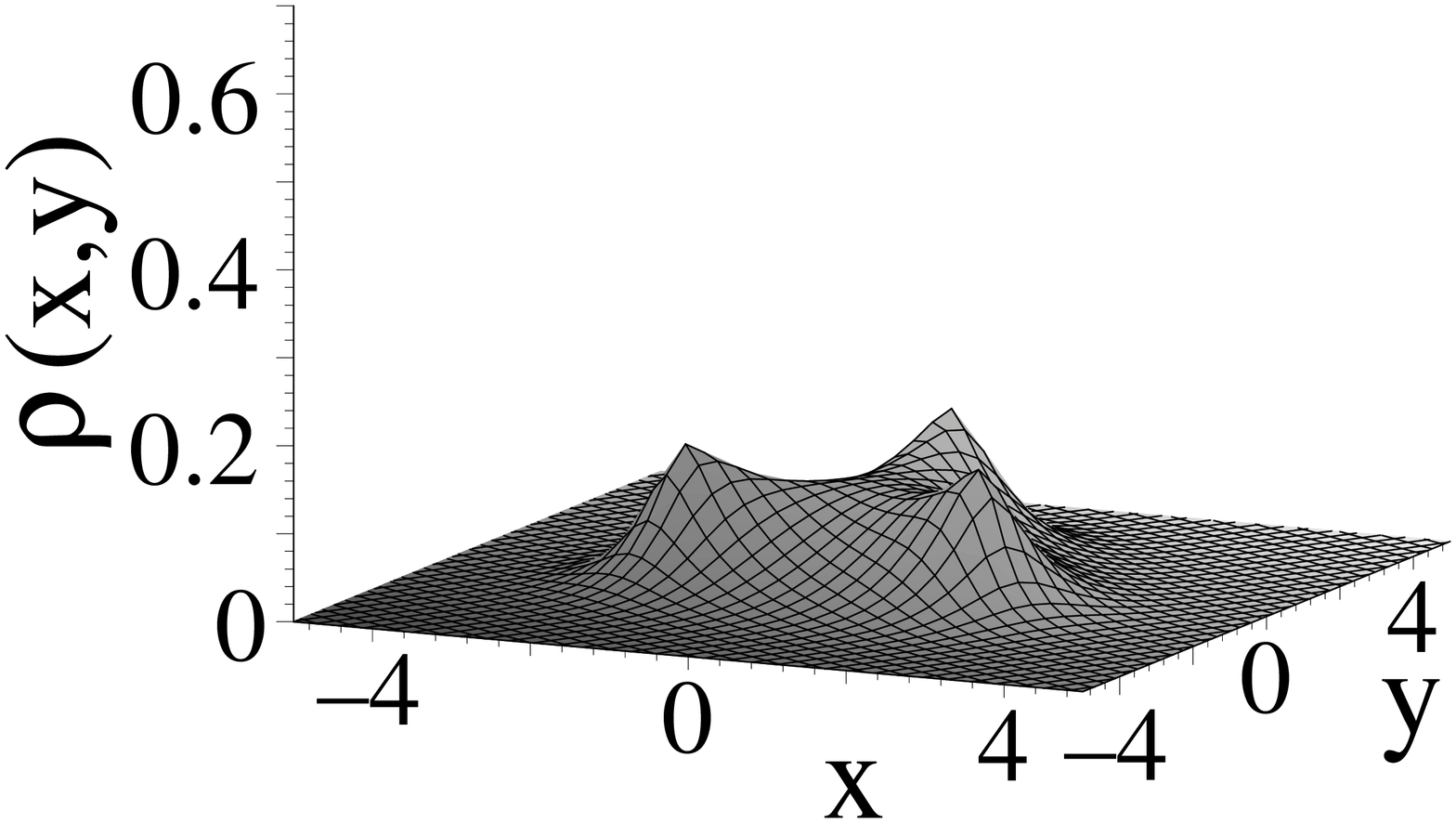}}
\put(120,100){$R=3.92$ a.u.}
\end{picture}
& &
\begin{picture}(200,150)(0,0)
\put(0,150){\includegraphics[height=120pt,width=120pt,angle=-90]{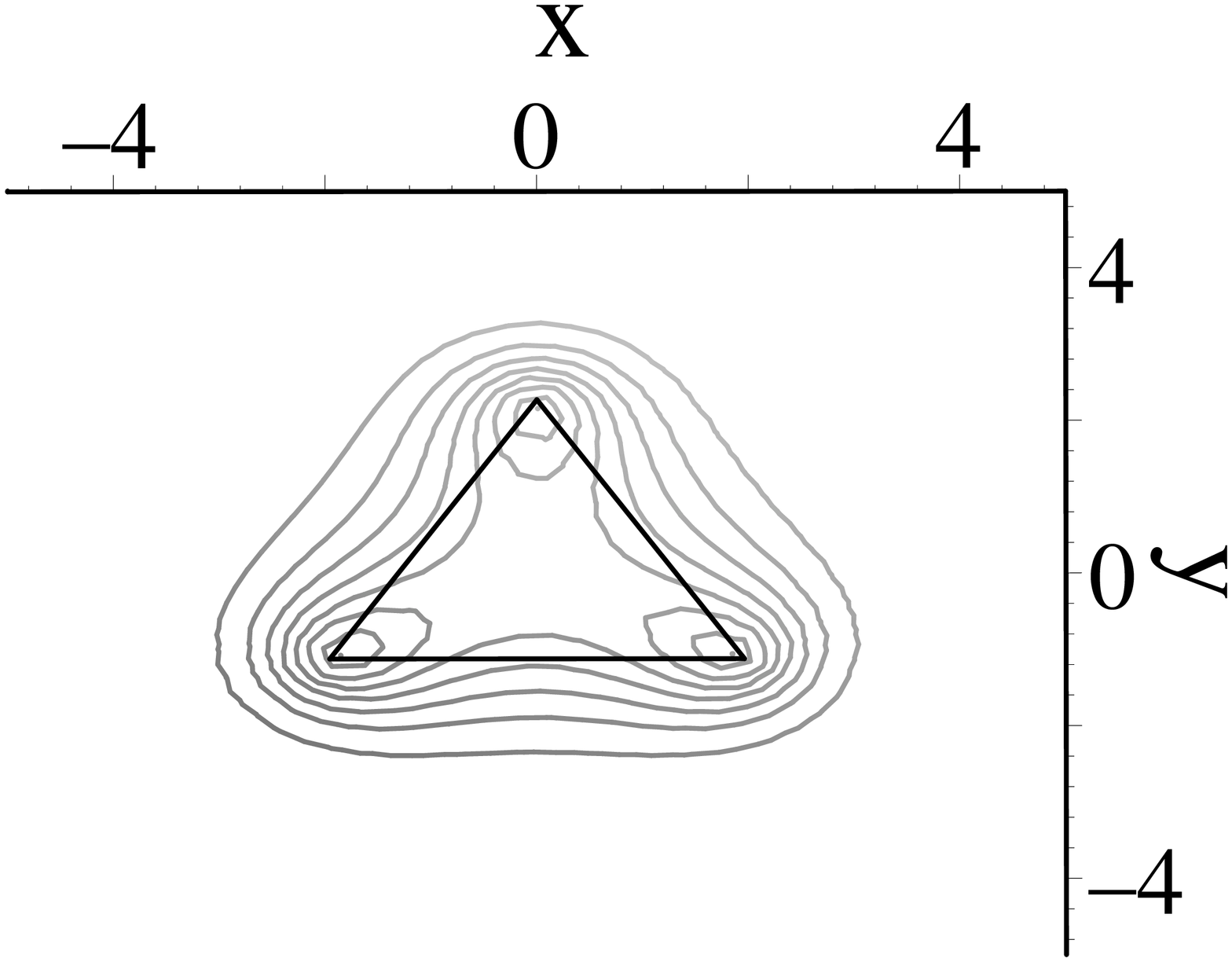}}
\end{picture}
\\[-10pt]
\begin{picture}(200,150)(0,0)
\put(0,150){\includegraphics[width=120pt,angle=-90]{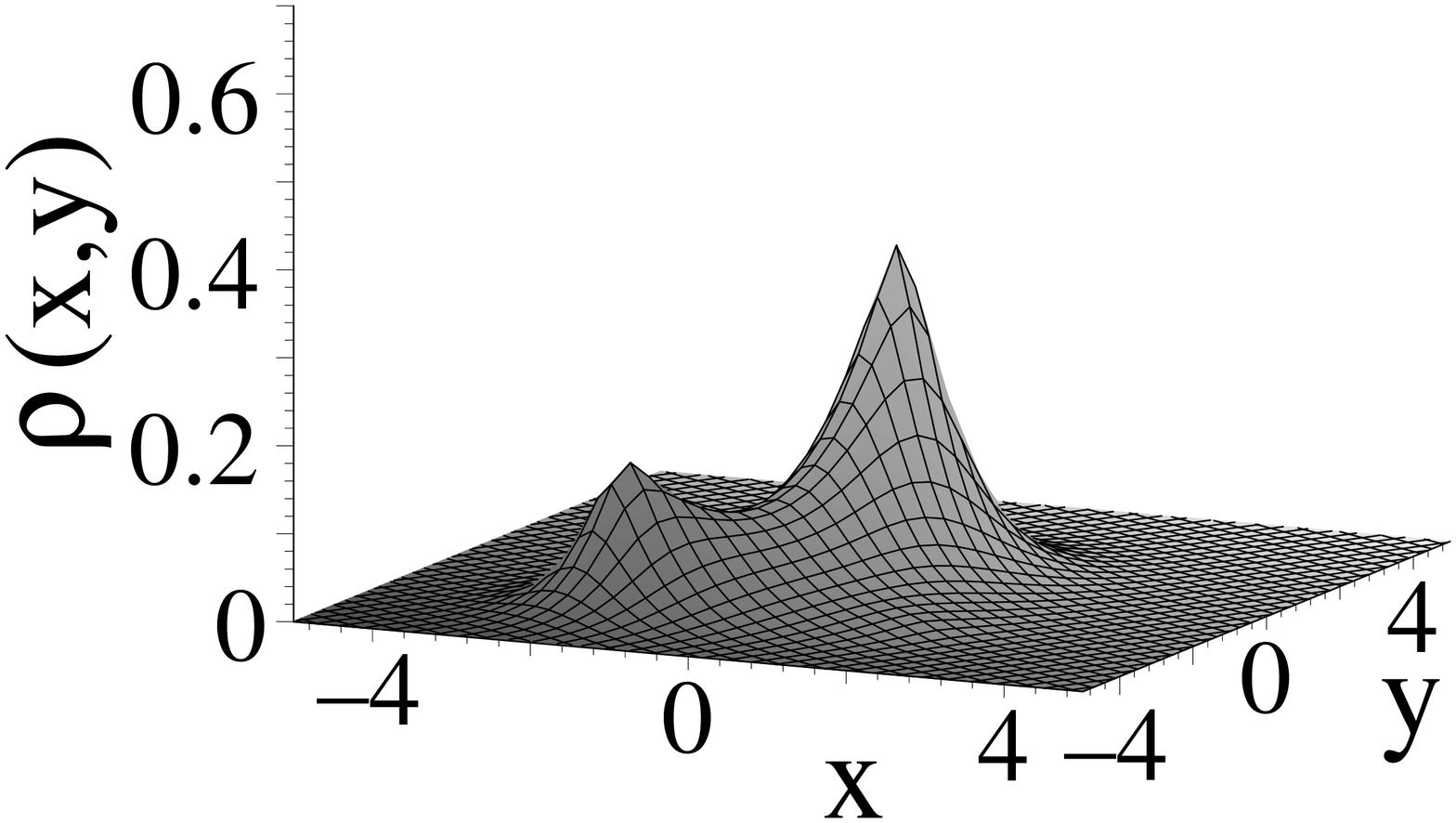}}
\put(120,100){$R=3.94$ a.u.}
\end{picture}
 & &
\begin{picture}(200,150)(0,0)
\put(0,150){\includegraphics[height=120pt,width=120pt,angle=-90]{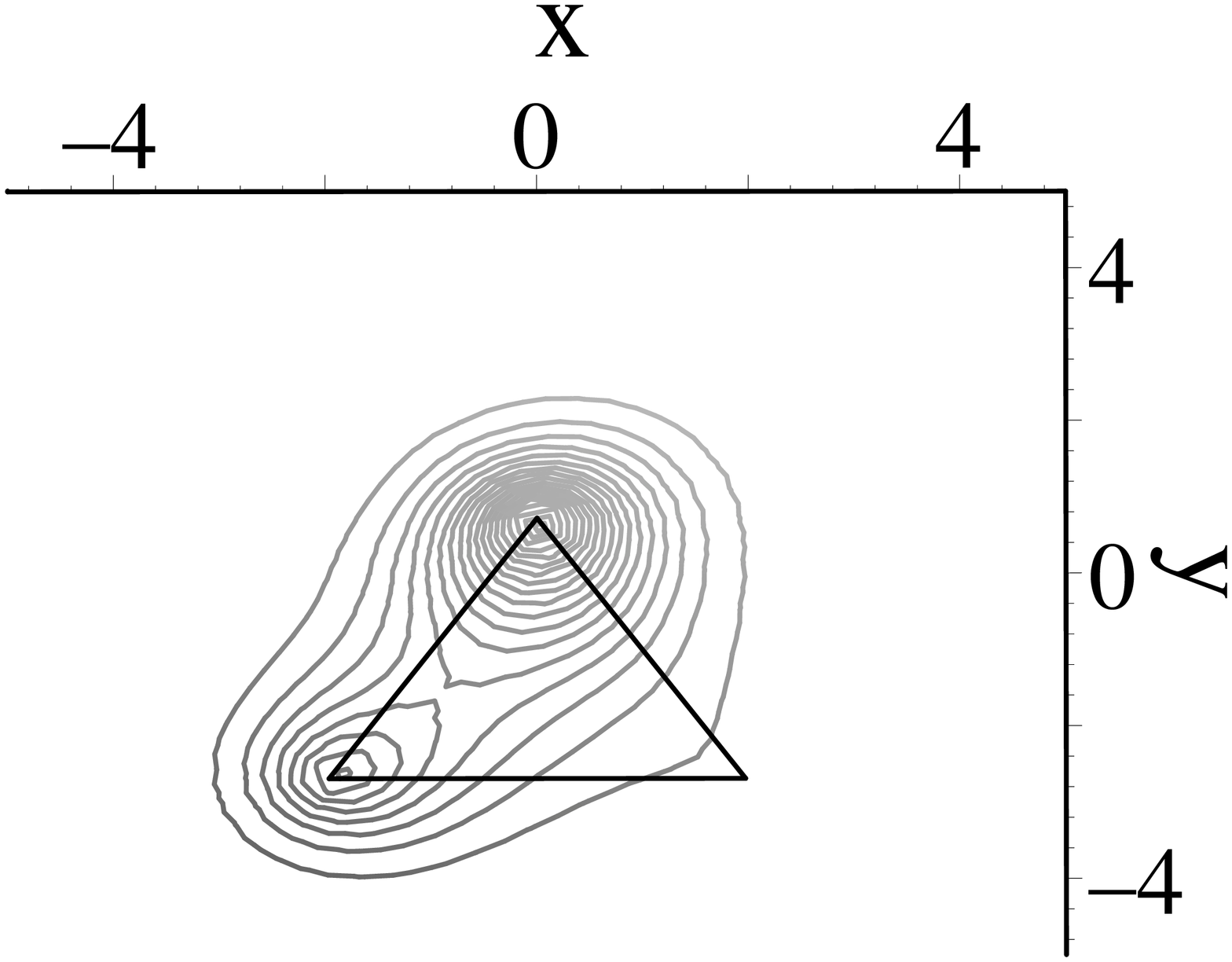}}
\end{picture}
\\[-10pt]
\begin{picture}(200,150)(0,0)
\put(0,150){\includegraphics[width=120pt,angle=-90]{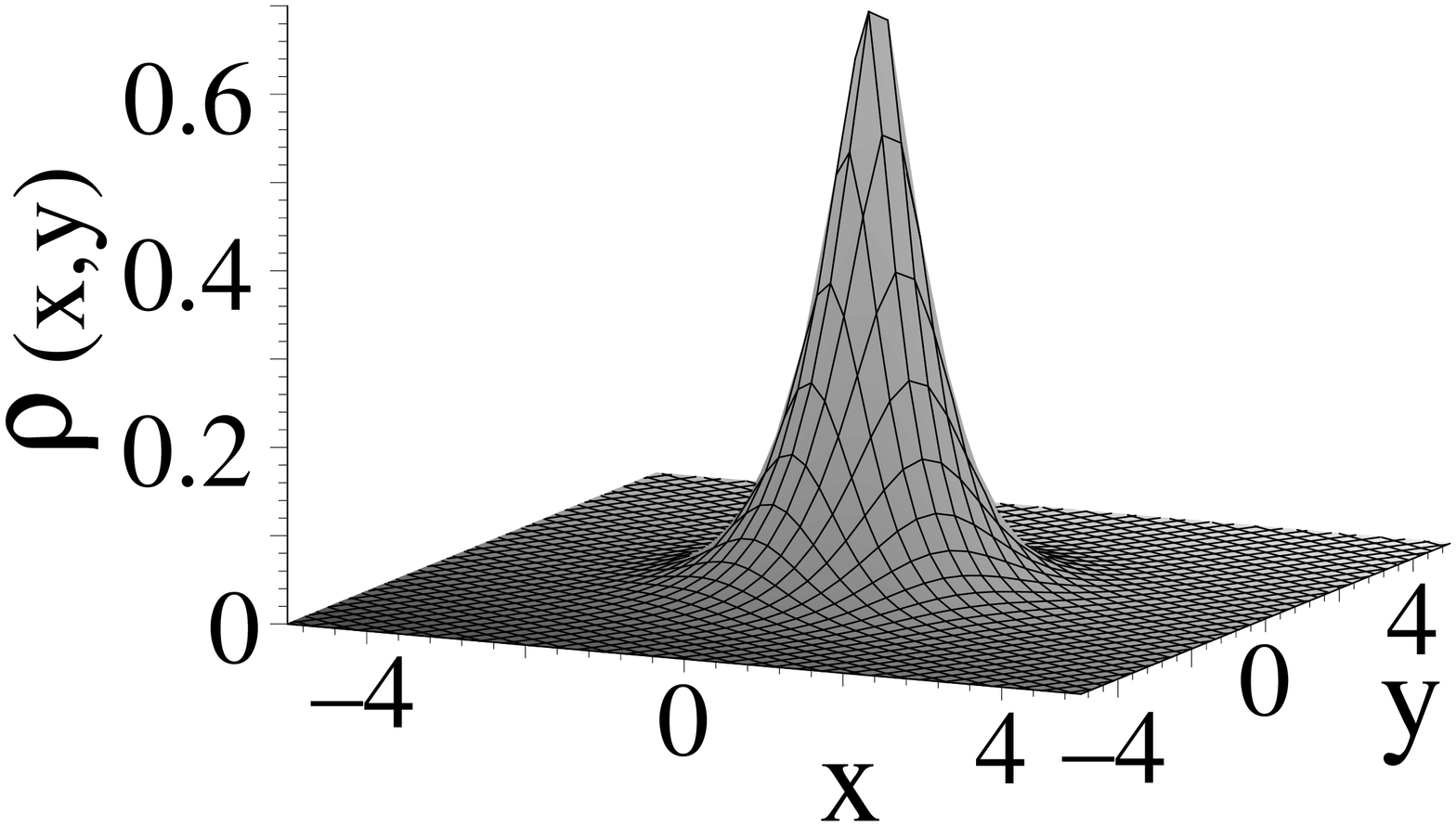}}
\put(120,100){$R=6.0$ a.u.}
\end{picture}
 & &
\begin{picture}(200,150)(0,0)
\put(0,150){\includegraphics[height=120pt,width=120pt,angle=-90]{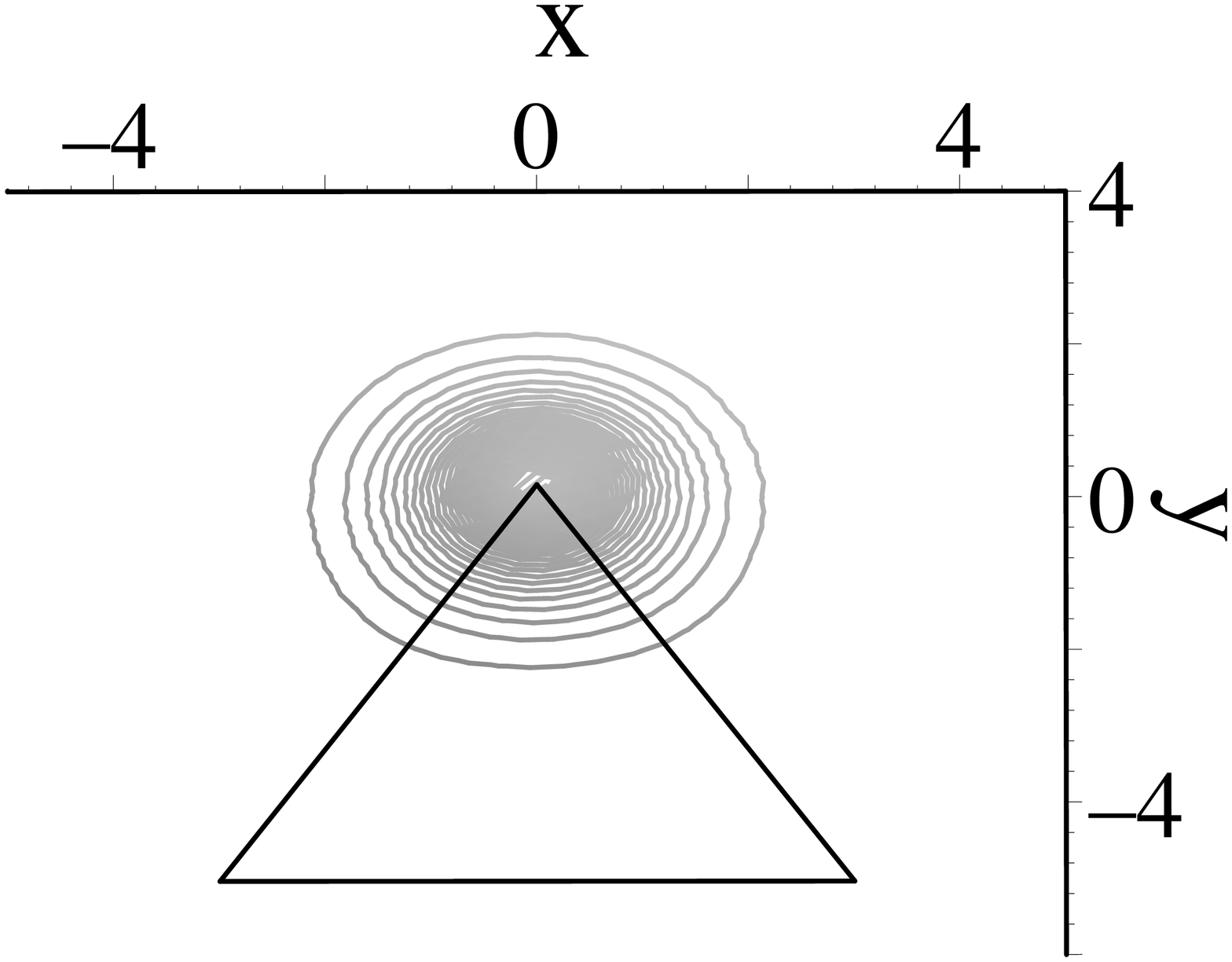}}
\end{picture}
\end{array}
\]
\caption{Evolution with $R$ of the integrated, normalized (to unity),
  electronic distributions
 $\rho(x,y) = \int |\Psi|^2(x,y,z) dz $ for $H_3^{++}$ in an
 equilateral triangular configuration at $B=10^9$G. The coordinates
 $x,y$ are given in a.u.}
   \label{fig:8}
\end{figure*}

\begin{figure*}
\[
\begin{array}{ccc}
\begin{picture}(200,150)(0,0)
\put(0,150){\includegraphics[width=120pt,angle=-90]{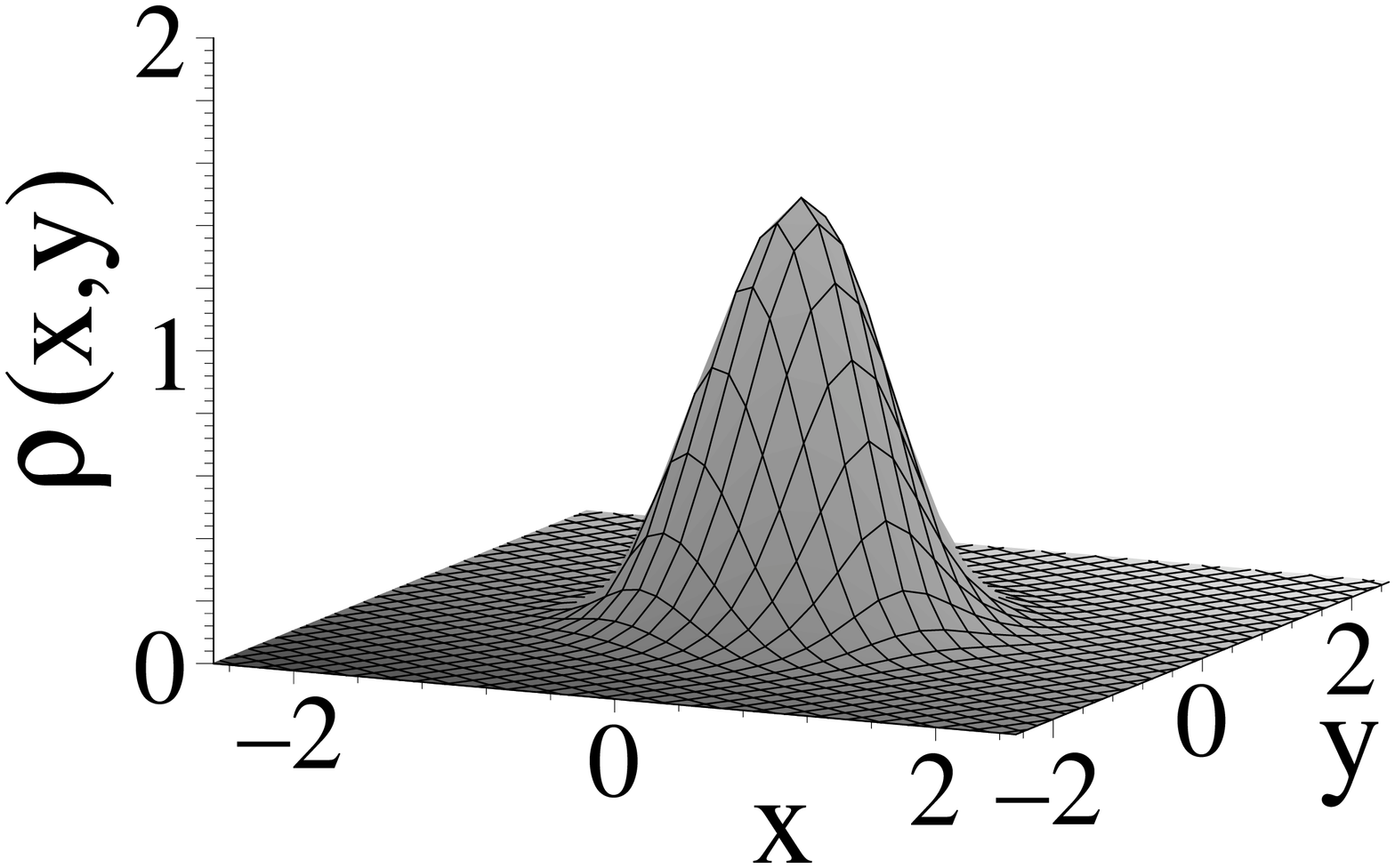}}
\put(120,100){$R=R_{eq}$}
\end{picture}
&  &
\begin{picture}(200,150)(0,0)
\put(0,150){\includegraphics[height=120pt,width=120pt,angle=-90]{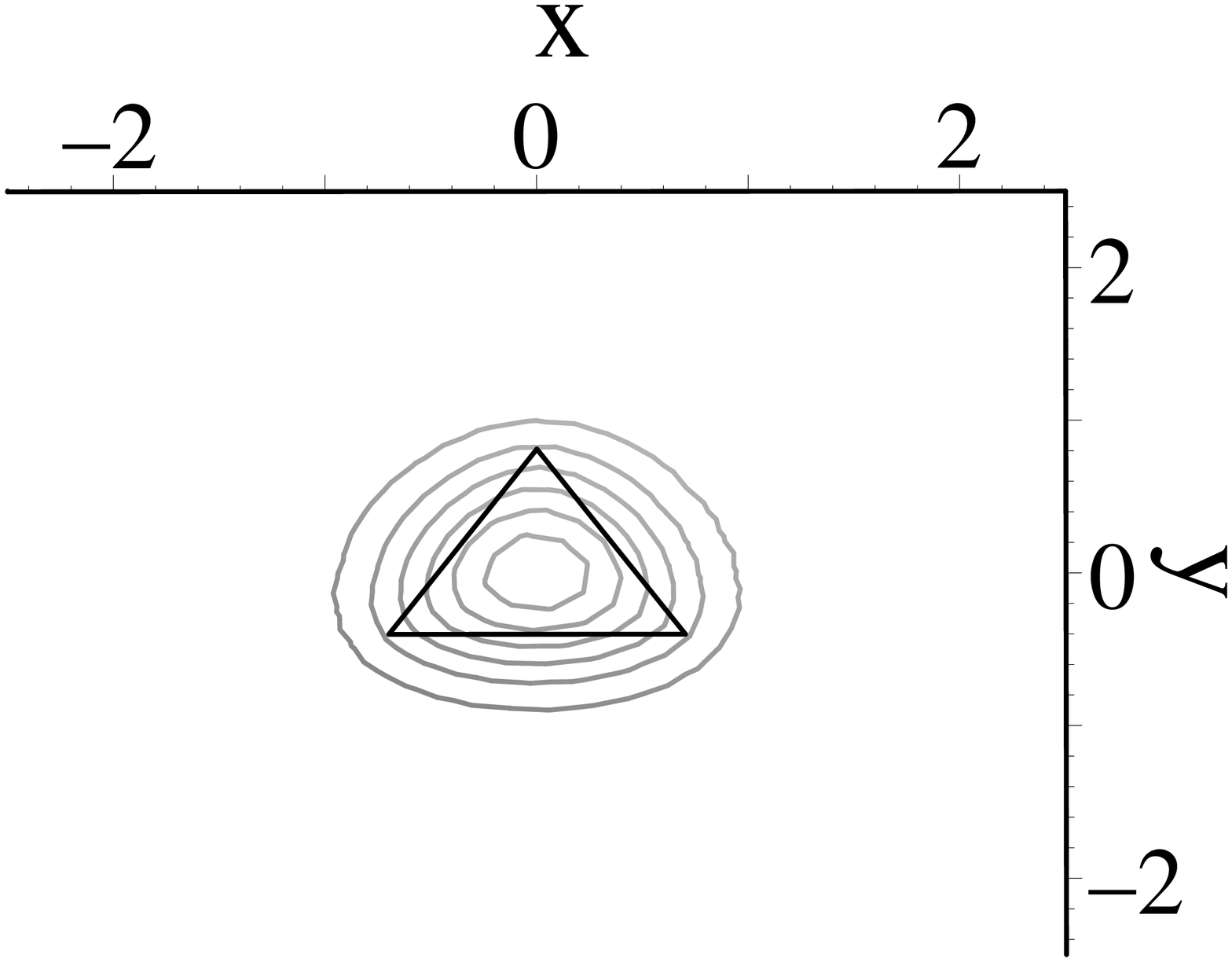}}
\end{picture}
\\[-10pt]
\begin{picture}(200,150)(0,0)
\put(0,150){\includegraphics[width=120pt,angle=-90]{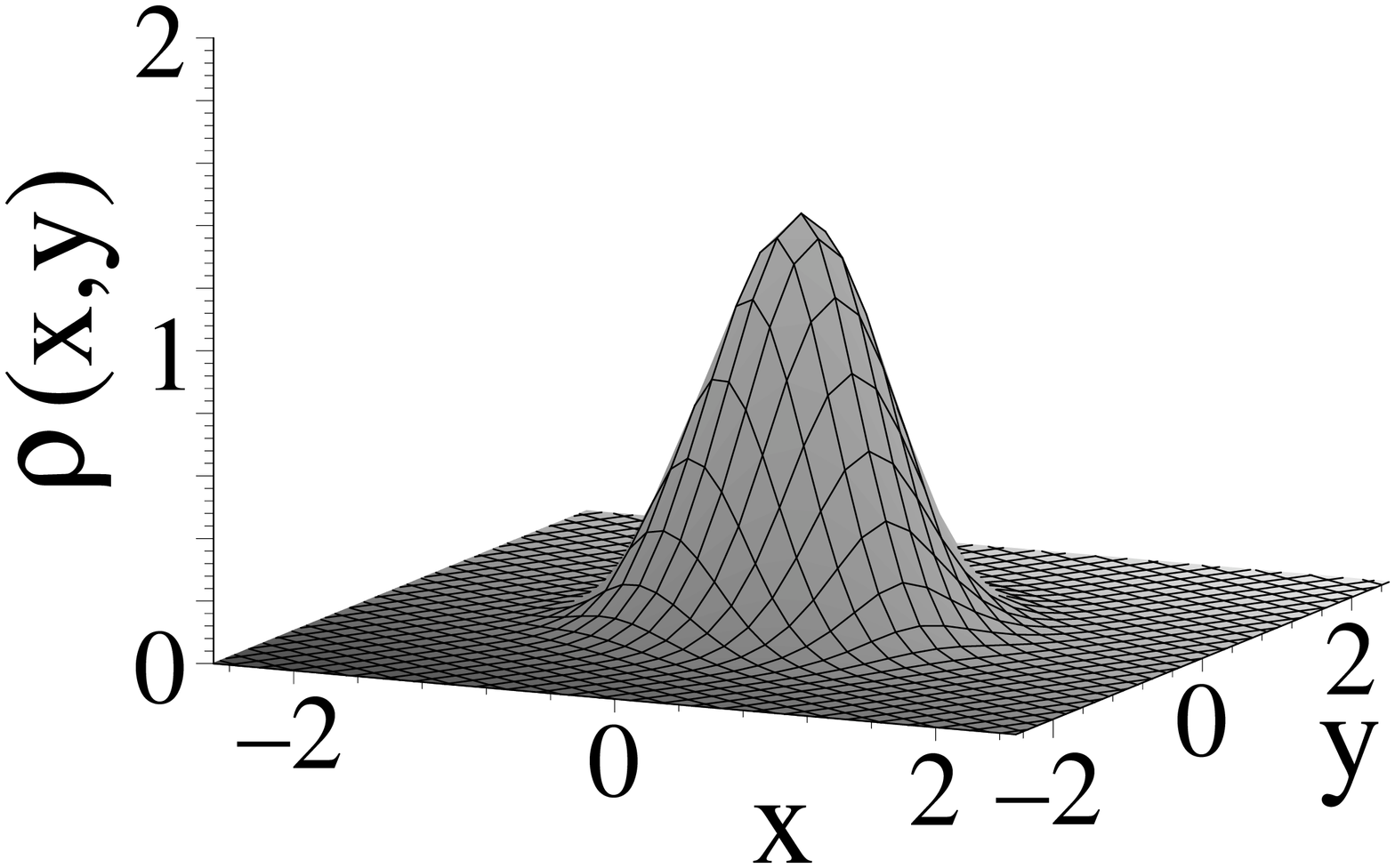}}
\put(120,100){$R=1.6$ a.u.}
\end{picture}
& &
\begin{picture}(200,150)(0,0)
\put(0,150){\includegraphics[height=120pt,width=120pt,angle=-90]{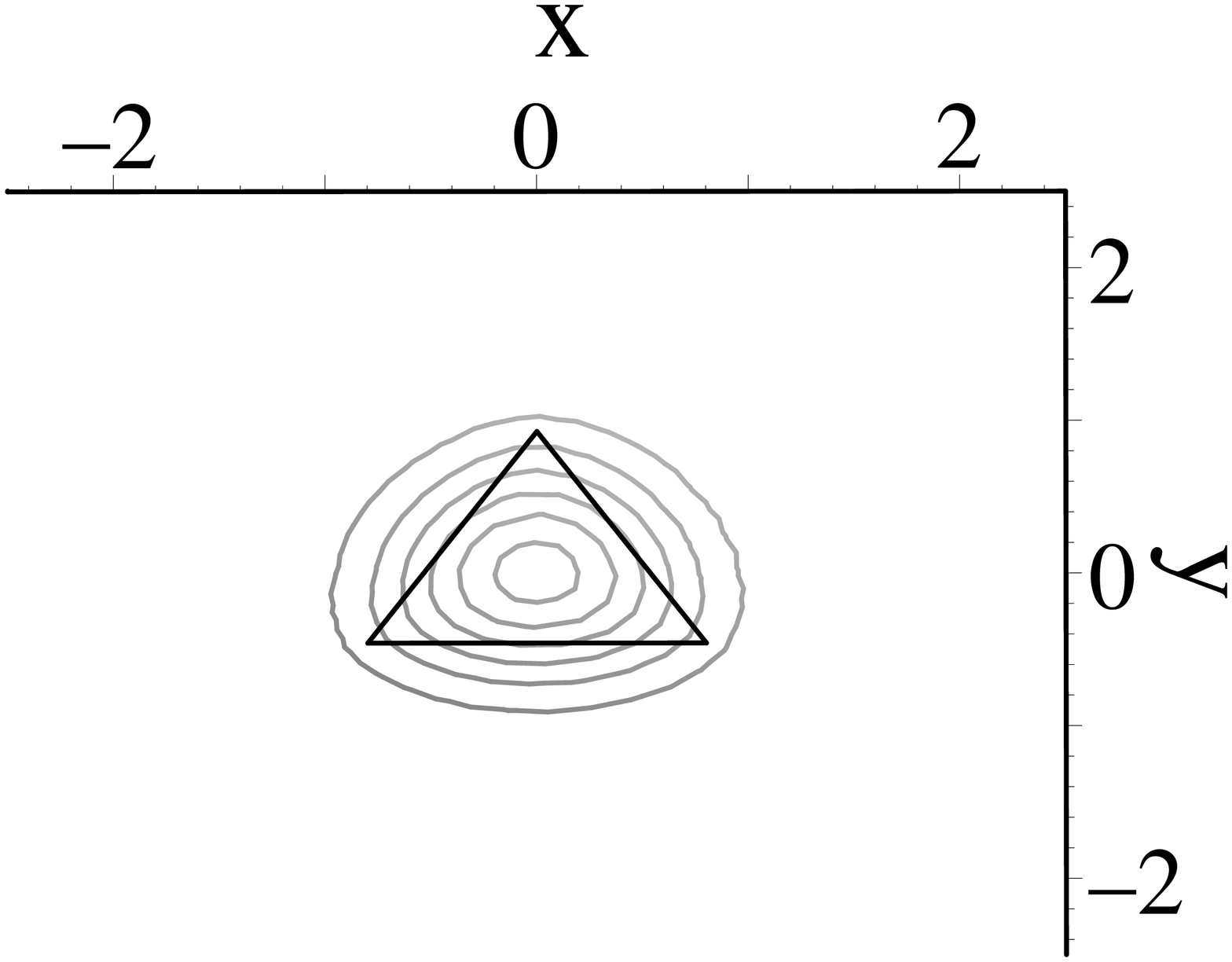}}
\end{picture}
\\[-10pt]
\begin{picture}(200,150)(0,0)
\put(0,150){\includegraphics[width=120pt,angle=-90]{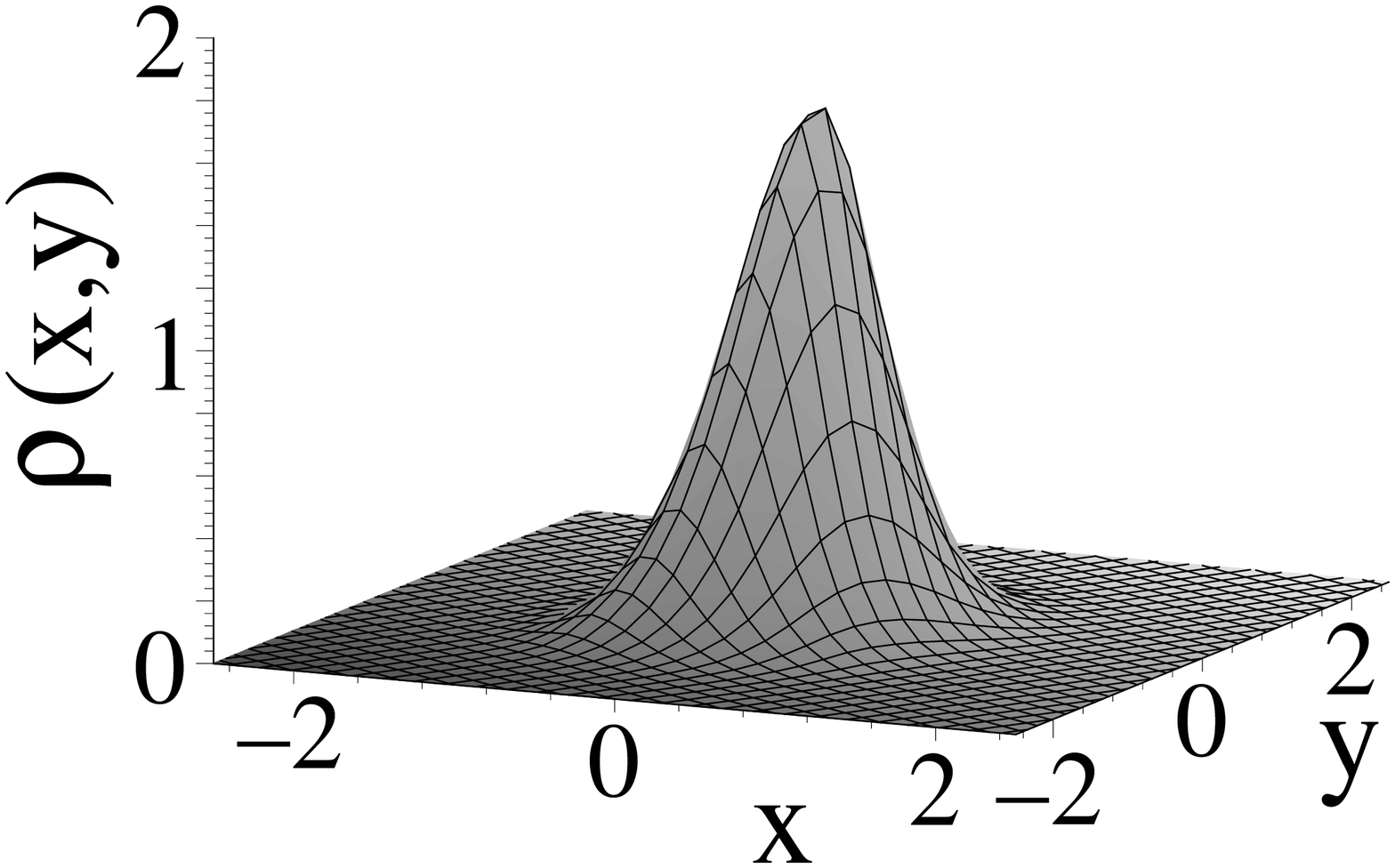}}
\put(120,100){$R=1.7$ a.u.}
\end{picture}
 & &
\begin{picture}(200,150)(0,0)
\put(0,150){\includegraphics[height=120pt,width=120pt,angle=-90]{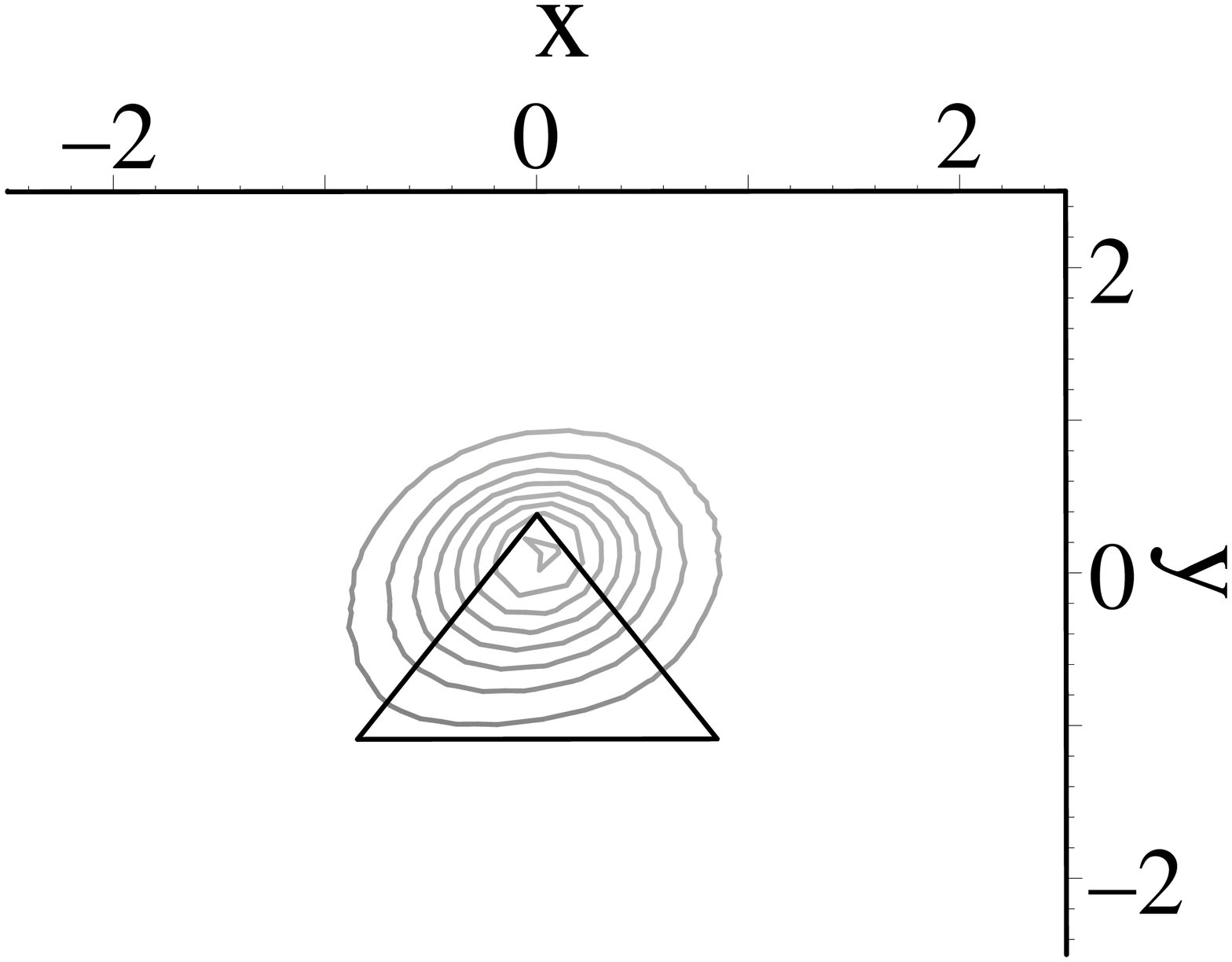}}
\end{picture}
\\[-10pt]
\begin{picture}(200,150)(0,0)
\put(0,150){\includegraphics[width=120pt,angle=-90]{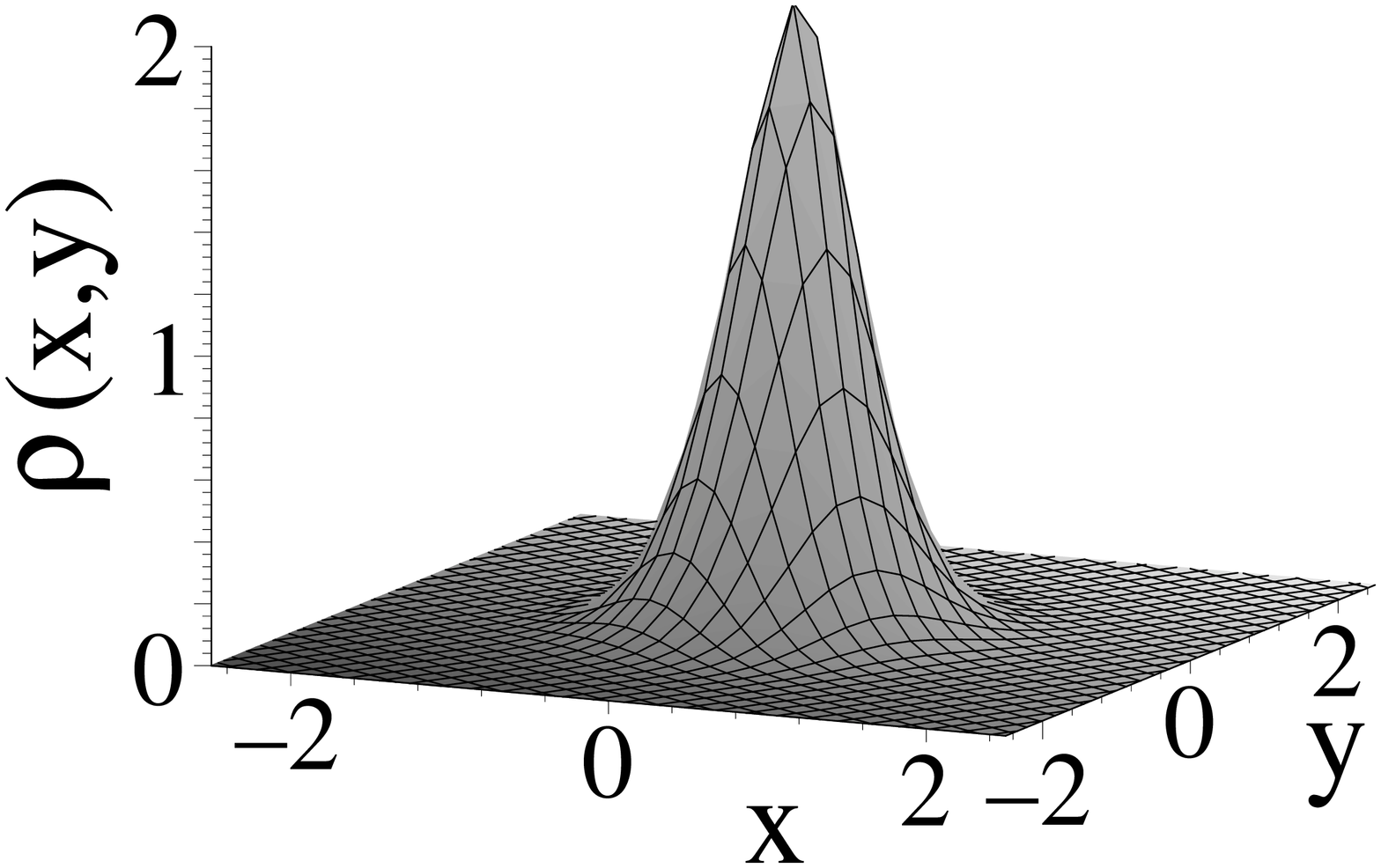}}
\put(120,100){$R=2.5$ a.u.}
\end{picture}
 & &
\begin{picture}(200,150)(0,0)
\put(0,150){\includegraphics[height=120pt,width=120pt,angle=-90]{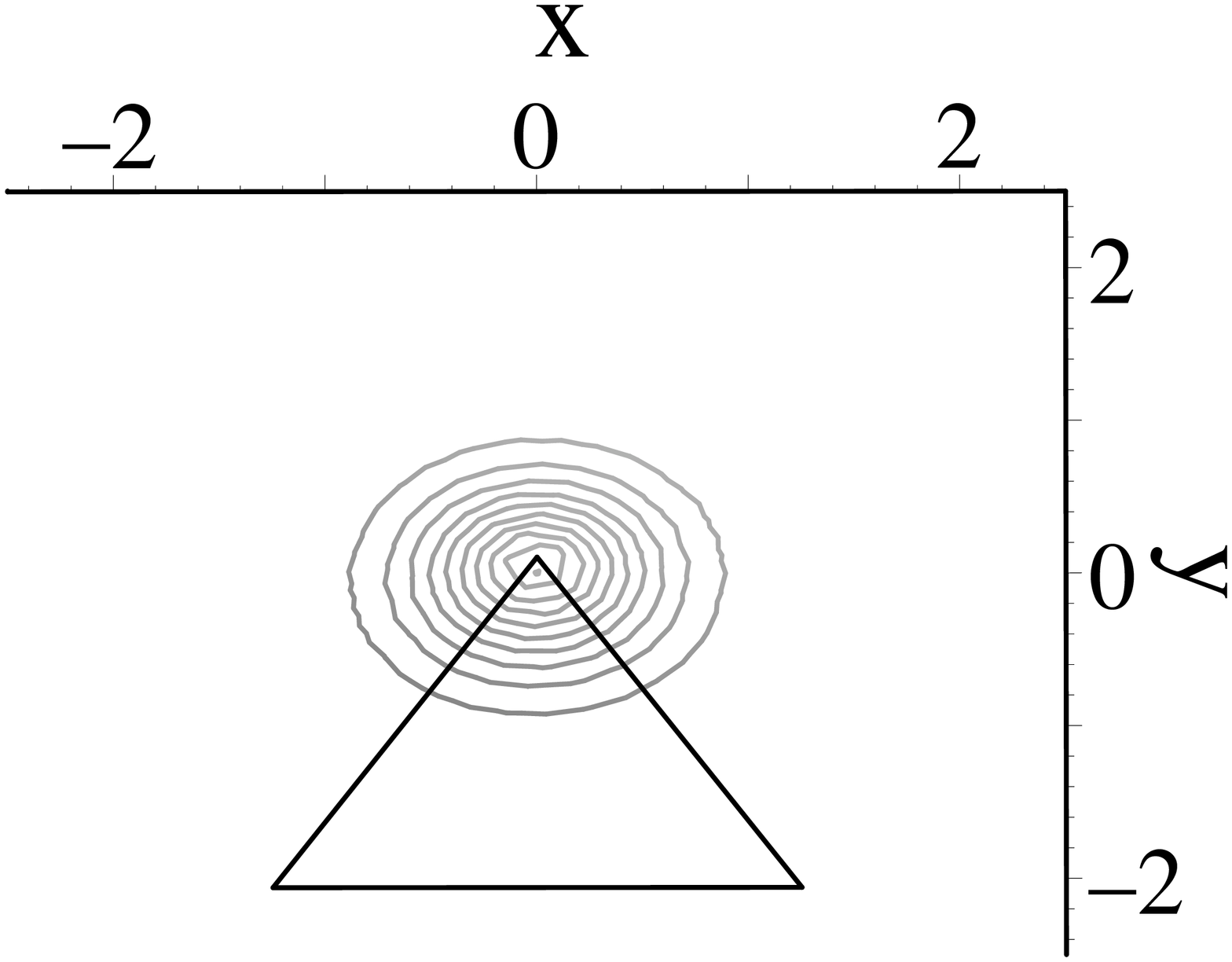}}
\end{picture}
\end{array}
\]
\caption{Evolution with $R$ of the integrated, normalized (to unity),
  electronic distributions
 $ \rho(x,y) = \int |\Psi|^2(x,y,z) dz $ for $H_3^{++}$ in an
 equilateral triangular configuration at $B=10^{10}$G.
 The coordinates $x,y$ are given in a.u.}
   \label{fig:9}
\end{figure*}

\begin{acknowledgments}
The authors wish to thank B.I.~Ivlev (IF-UASLP, Mexico) for useful
conversations.

This work was supported in part by DGAPA Grant \# IN120199 (M\'exico).
\end{acknowledgments}




\begin{thebibliography}{99}

\bibitem{Lopez:2000}
        J.C.~L\'opez~Vieyra,
        {\it Rev.Mex.Fis. \bf 46}, 309 (2000)

\bibitem{Turbiner:1999}
        A.V.~Turbiner, J.C.~L\'opez~Vieyra, U.~Solis H.\\
        {\it Pis'ma v ZhETF \bf 69}, 800-805 (1999)\\
        {\it  JETP Letters \bf 69}, 844-850 (1999)
        (English Translation),  (\eprint{astro-ph/9809298})

\bibitem{Lopez-Tur:2000}
        J.C.~L\'opez~Vieyra and A.V.~Turbiner,
        {\it Phys.Rev. \bf A62}, 022510 (2000)

\bibitem{Larsen}
        D.~Larsen, {\it Phys.Rev. \bf A25}, 1295 (1982)

\bibitem{LL}
        L.~D.~Landau and E.~M.~Lifshitz, {\it Quantum
        Mechanics}, Pergamon Press (London) 1977

\bibitem{Schmelcher}
        P.~Schmelcher, L.S.~Cederbaum and U.~Kappes,\\
        in {\it Conceptual Trends in Quantum Chemistry}, 1-51,
        Kluwer Academic Publishers, Dordrecht (1994)

\bibitem{Tur}
        A.V.~Turbiner,
        {\it  ZhETF \bf 79}, 1719 (1980),\\
        {\it Soviet Phys.-JETP \bf 52}, 868-876 (1980) (English
        Translation);

        {\it Usp. Fiz. Nauk. \bf 144}, 35 (1984),\\
        {\it Sov. Phys. Uspekhi \bf 27}, 668 (1984)
        (English Translation);

\bibitem{Lopez:1997}
        J.C.~L\'opez~Vieyra, P.O.~Hess, A.V.~Turbiner,
        {\it Phys.Rev. \bf A56}, 4496 (1997),
        (\eprint{astro-ph/9707050})


\bibitem{Potekhin:2001}
        A.~Potekhin, A.~Turbiner,
        {\it Phys.Rev. \bf A63}, 065402 (2001) 1-4,
        (\eprint{physics/0101050})

\end{thebibliography}
\end{document}